\documentclass[12pt,a4paper,twoside]{article}

\usepackage{graphicx}
\usepackage{epic}
\usepackage{epstopdf}
\usepackage{amssymb}
\usepackage{amsmath}
\usepackage{sectsty}

\usepackage[utf8x]{inputenc} 
\usepackage[english]{babel}

\sectionfont{\huge}

\begin{document}

\begin{center}
{\huge\bf On Mathematics of Bubbles}\\[1ex]
{\huge\bf in Dynamical Systems}
\end{center}

\vspace{0.3cm}
\vspace{0.0cm}
\begin{center}
{\it Petr Kl\'{a}n, Department of Software Engineering, Faculty of Information Technology, Czech\\[1ex] Technical University in Prague, petr.klan@fit.cvut.cz}
\end{center}

%\vspace{0.7cm}
%\noindent
%{\bf Abstract:} A new concept called biased derivative is proposed. It has a potential to better understand and model some aspets of dynamical systems such as activator and inhibitor functioning.

\vspace{1cm}

\section{Introduction}

\vspace{0.3cm}\noindent Dynamical systems are used as a general label to discuss issues regarding dynamics in all environments where changes take place from the simplest particles to social system \cite{ber}. There are much effort to introduce dynamical systems and their activating and inhibiting mechanisms more formally \cite{sch}. It results in a preference of some mathematical models and less sure about others. Furthermore, the specific environments apply a particular approach \cite{kah, mei, sti}.\\
\\ 
In this paper, a mathematical concept is introduced in order to model anomalies in the behavior of dynamical systems. For this purpose, a specific kind of derivative (called biased derivative) is introduced and used. Motivation for using this kind of derivative is given by inherent part of these systems: the tendency to establish patterns incorporating overshoots or bubbles.\\
\\
The paper is organized as follows. Section 2 explains, how the mathematical model method works. Section 3 presents a few illustration examples for being motivated in principles of the dynamical system modelling.  In Section 4, biased derivative as a specific tool inspired by anomalous behavior of dynamical systems is introduced. Mathematics of dynamical system models is proposed in Section 5. Section 6 has an illustrative character given by the simple case study of bubble formation in a dynamical system. 

\section{The Model Method}

\begin{figure}[htb] 
\begin{center}
\setlength{\unitlength}{0.8mm}
\begin{picture}(140,40)
%Control blocks
\thicklines 
\put(60,10){\framebox(25,20){\bf System}}

\thinlines
\put(20,20){\vector(1,0){40}}
\put(85,20){\vector(1,0){40}}

\put(30,22){\makebox(0,0){Input}}
\put(110,22){\makebox(0,0){Output}}
\end{picture}
\caption{A single system block.} \label{block}
\end{center}
\end{figure}
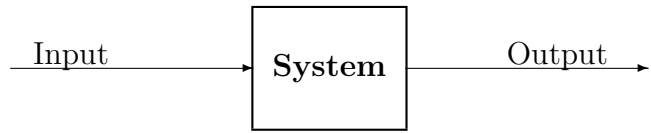

\vspace{0.3cm}\noindent  Dynamical systems can be represented by single blocks having input and output as shown in Fig. \ref{block}. They process an stimulus from the related environment to generate some response affecting this environment. Systems carry out activities that are, in many aspects, response--intensive; and in some complex environments such as societies they manifest many features and actions that are strange or difficult to predict.

\vspace{0.3cm}\noindent It has always been the purpose of science to make models \cite{ric}. A mathematical theory which seeks to explain and to predict the events always deals with a simplified model, a mathematical model in which only things pertinent to the behavior under consideration enter \cite{pie}.\\ 
\\
Dynamical systems are considered to be mathematically modelled. The latter enables to more understand their behaviour and to predict what their action will be, if submitted to a known input stimulus. The simplest method for representing the behaviour of a dynamical system is to mimic it with a mathematical description in the form of a differential equation, for example: $T\dot{y}+y=Ku$, where $u,y$ are system input and output, respectively, and $K,T$ are the significant structural parameters.
\\
\\
At beginning, the modelled system is excited to learn its behaviour. Immediately, two behaviours will be compared, that of the real system and that of the mathematical representation, the model. If the behaviour of the real system and the model is equal, for example, responses to the known input stimulus are the same, it is possible to state that a model of the real system is identified. Note that this procedure is indeed that of experimental science advocated by G. Galilei providing a closed loop experimental discovery opposed to the deductive path of mathematics. Unfortunately, real systems are not so simple to totally encompass them by a mathematical description. Therefore, models mostly tend towards a similarity of behaviour with real systems being represented.   

\section{On Models of Dynamical Systems}

\begin{figure}[htbp] \begin{center}
\setlength{\unitlength}{0.8mm}\begin{picture}(120,104)
%Single block
\thicklines \put(35,10){\framebox(40,20){\bf Inhibitor}}\put(35,50){\framebox(40,20){\bf Activator}}
%Lines
\thinlines \put(75,60){\line(1,0){20}}\put(95,60){\line(0,1){35}}
\put(95,60){\line(0,-1){40}}\put(95,95){\line(-1,0){90}}
\put(95,60){\circle*{0.9}}

\put(95,20){\vector(-1,0){20}}
\put(5,60){\circle{10}}\put(10,60){\vector(1,0){25}}
\put(5,95){\vector(0,-1){30}}\put(35,20){\line(-1,0){30}}
\put(5,20){\vector(0,1){35}}

\put(2,52){\makebox(0,0){$-$}}
\put(2,68){\makebox(0,0){$+$}}
\end{picture} \caption{\small An inner activator--inhibitor structure of the dynamical system.} \label{dynsys}
\end{center}
\end{figure}
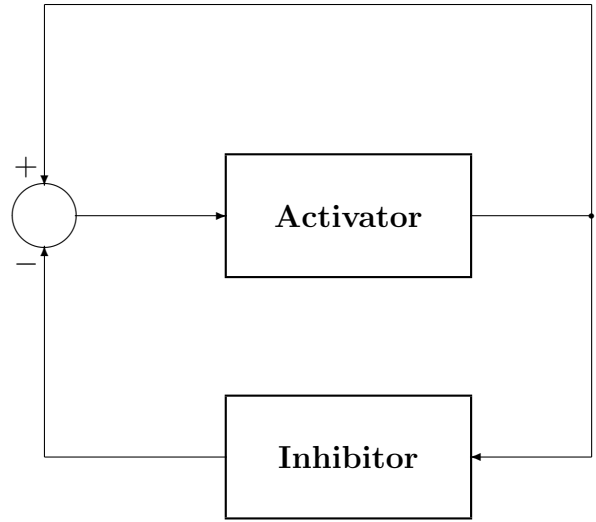

\vspace{0.3cm}\noindent Our experience indicates that the behavior of dynamical systems is neither as determined as that of the Pythagorean theorem nor as simply random as the throw of a die or as the drawing of balls from a mixture of black and white balls \cite{pie}.\\
\\
Due to their response--intensive activities, dynamical systems have an inherent tendency to evolve over time \cite{she}. When growing, dynamic systems self--enhance the initial deviation from the mean. Otput of dynamical systems grows since attracted by input stimulus similarly as cities grow since they attract more people. Based on size of input values, self--enhancing processes evoke inhibiting reactions similarly as the increasing noise and traffic may discourage people from moving into a growing city. Furthermore, if the inhibition follows with some delay, the activated self--enhancing reaction can cause an oveshoot.\\
\\
Imagine the functioning of a dynamical system by the sand dune paradox. Naively, one would expect that the wind in the desert causes a structureless distribution of the sand. However, wind, sand and surface structure together represent an unstable system where dunes are formed. Sand accumulates behind the wind shelter. Dune begins to grow increasing the wind shelter which self--enhances the deposition of sand. But the sand, once settled in the dune, cannot participate in dune formation. Hence, the inhibiting reaction results from removal of sand able to participate in dune grow. In this way, the growth of dunes is reduced. This leads to self--regulated dune patterns \cite{mei}.\\
\\
The model scheme of a dynamical system is considered to be composed of an activator and an inhibitor part as shown in Fig. \ref{dynsys}. The activator represents a self--enhancing substance of the system whilst the inhibitor relates to the inhibiting activity. The development of both parts participates in a steady state. Notice that the stable behavior requires some specific inhibiting reactions.   

\section{Biased Derivative}

\vspace{0.3cm}\noindent Describing mathematical dynamical systems include a function (i.e. a rule) which tells, given the current output, what the output of the system will be in the next instant of time. Typically, it 
is provided by equations which contain derivatives such as $T\dot{y}+y=Ku$. Here, derivative is defined by the ratio $\dot{y}=[y(t+\Delta t)-y(t)]/\Delta t$ for an infinitesimally small $\Delta t$. I.\,Newton called it  fluxion.\\
\\
Dynamical systems are associated with entities such as concatenation, wide spreading, noise, disturbances, delays \ldots concerning a piece of complexity and irregularity to actual values of $y(t)$. The latter becomes biased. It reflects in a {\it biased derivative}, algebraically

\begin{equation}
y^\odot = [y(t+\Delta t)-(1-\epsilon \Delta t)y(t)]/\Delta t
\end{equation}
where $\epsilon$ is taken as a coefficient given by the irregular conditions \cite{kla}. It results in a bias of the derivative. If

\begin{itemize}
\item positive, then the derivative can tend to be overestimated,
\item negative, then the derivative can tend to be underestimated, 
\item zero, then biased derivative becomes the ordinary one, 
\item proportional to $y(t)$, then the derivative can tend to create bubbles, 
\item proportional to $y(t)^2$, then the derivative can tend to have a chaotic behavior.
\end{itemize}
Therefore, the coefficient has a potential to categorize behavior of derivatives and thus the behavior of  dynamical systems associated with these derivatives. 

\section{Biased Dynamical Systems}

\vspace{0.3cm}\noindent A determining property of the biased derivative is given by the following theorem.
\\
\\
{\bf Theorem:} If an ordinary derivative $\dot{y}$ is known, then the biased derivative $y^\odot$ is determined by 

\begin{equation}
y^\odot =\dot{y}+\epsilon y(t). \label{bides}
\end{equation} 
Proof of this rule is given by a simple rearrangement of (1). By the method of (2), differential equation  
$T\dot{y}+y=Ku$ is represented in the form of $y^\odot=Ku/T$, where $\epsilon=1/T$. It well illustrates the concept of biased derivative since response to a unit step input is $y(t)=K(1-e^{-t/T})$ for the zero initial condition $y(0)=0$ and $t\geq 0$. It results in a self--regulating system in which inhibiting mechanism is automatically included.\\
\\
Concept of biased derivative, among others, covers self--regulating scheme of logistic equations and predator--prey interactions. Logistic equations are the most popular models for the concept of saturation in grow of population $N(t)$ with the carrying capacity $K$ and multiplicative factor $\sigma$. They are expressed by $\dot{N}=\sigma N (1-N/K)$. The latter becomes $N^\odot=-\sigma N^2/K$ by a use of the biased derivative, where $\epsilon=-\sigma$.  
\\
\\
Similarly, consider a set of coupled double population $N_1,N_2$ equations of the form (Kahn, 1990) $\dot{N_1}=\epsilon_1 N_1-\gamma_1 N_1 N_2$, $\dot{N_2}=-\epsilon_2 N_2+\gamma_2 N_1 N_2$, where $\epsilon_1,\epsilon_2,\gamma_1,\gamma_2$ are multiplicative factors. With a use of biased derivatives it follows that $N_1^\odot=-\gamma_1 N_1 N_2$ and $N_2^\odot=\gamma_2 N_1 N_2$, where $\epsilon=-\epsilon_1$ (first equation) and $\epsilon=\epsilon_2$ (second equation).\\
\\
Arrange the activating and inhibiting paths according to their effects and combine these factors to an input--output relationship based on the biased derivative

\begin{equation}
y^\odot=f(u)
\end{equation}
where $f$ denotes a function of the input.  In a single linear form, the model becomes $y^\odot=Ku$, where the constant $K$ plays a role of the system gain. A more complex form of dynamical systems is expressed by $y^\odot=f(y,u)$ where the function $f$ includes two independent variables. Based on type of $\epsilon$, it improves to categorize behavior of dynamical system.  

\section{A Case Study of Bubbles}

\begin{figure}[htbp]
\begin{center}
\includegraphics[scale=0.6]{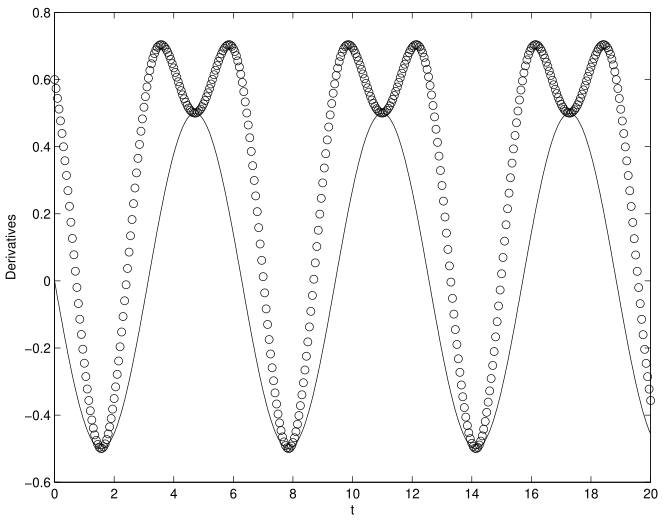}
\end{center}
\caption{\small Biased derivatives of harmonic signal: model -- solid, bubble -- o.}
\label{bbsig}
\end{figure}

\begin{figure}[htbp]
\begin{center}
\includegraphics[scale=0.2]{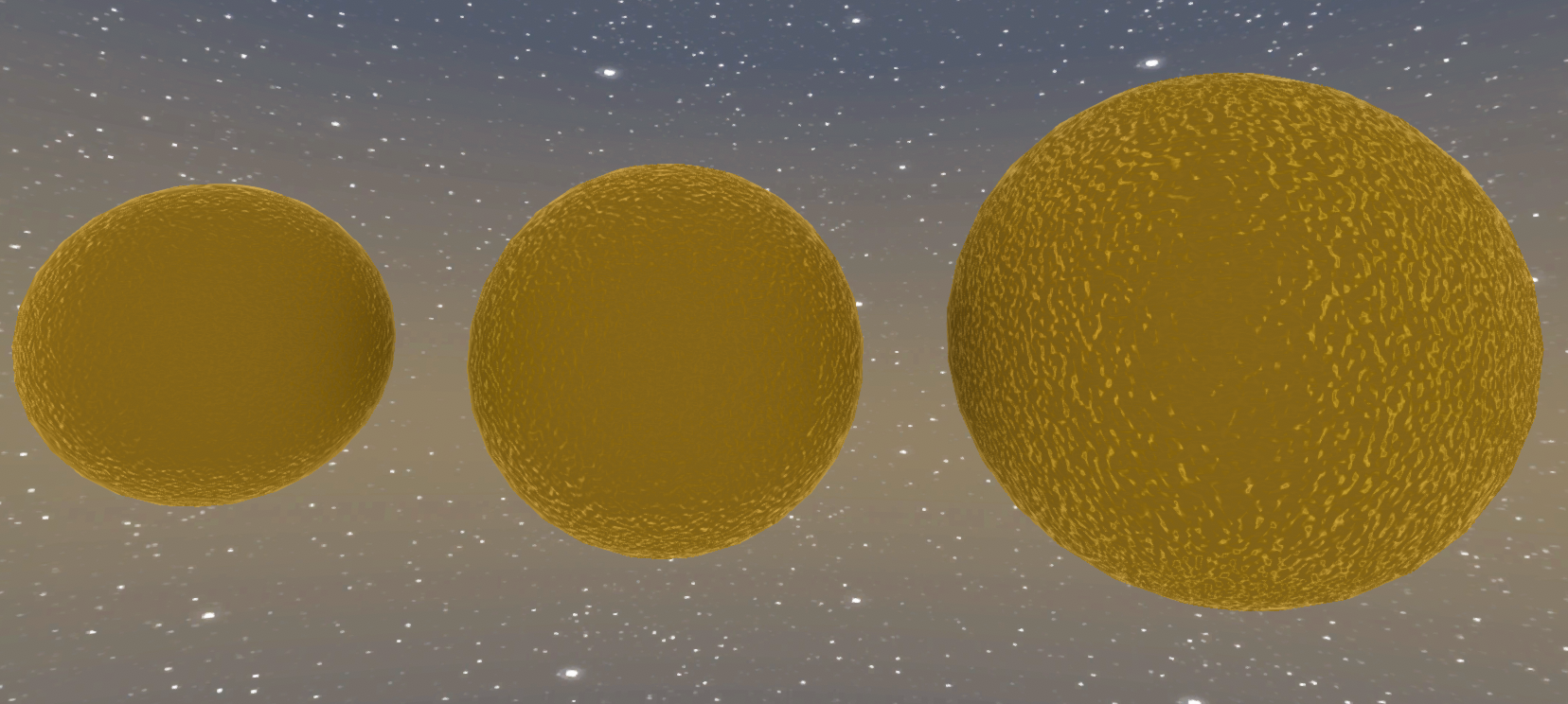}
\end{center}
\caption{\small Accumulation of changes in positive periods: left volume -- unbiased, middle volume -- constant bias, right volume -- variable bias. Modelled in social virtual platform Neos VR.}
\label{bbs}
\end{figure}

\vspace{0.3cm}\noindent  Naturally, every society creates periodic chains of variable length, which are difficult to predict. Biased derivatives are aimed at indexing society processes that are inherently unpredictable. Their indexing (or simply determining nature of $\epsilon$) helps to measure effects of rational/irrational decisions in order to model movement of societies. It is based on a wide scare that can be associated with irrational activity in a subjective measure, for example.\\ 
\\
Suppose a periodic behavior of some quantity in the society which will be identified with harmonic signal of the form $y(t)=\cos (t/2)$, where $t$ denotes time. This quantity accumulates according to changes given by biased derivatives in (\ref{bides}):

\begin{enumerate}
\item Bias free $\epsilon=0$ (model state).
\item Constant bias $\epsilon=0.6$ (proportional state).
\item Variable bias $\epsilon=0.6y(t)$ (bubble state).
\end{enumerate}
A comparative study of these states is shown in Fig.\ref{bbsig}. Here, the model state is represented by solid line and the bubble state by o line. It can be observed, that o line has tendency to create bubbles in its maximum area.\\
\\
Suppose the accumulation of changes in positive periods as represented by sizes of spheres in Fig. \ref{bbs}. Whilst the model state is smallest (left), the biased spheres tend to oversize this model. They  indicate two dynamical models of the form (\ref{bides}), first with $\epsilon=0.6$ (middle sphere) and the second model with $\epsilon=0.6y(t)$ (right sphere). The latter maximum volume can reflect accumulation of irrational activities based on a wide misinformation effect in society, for example.   

\section{Conclusion}

\vspace{0.3cm}\noindent In this paper a specific kind of derivative is used to describe specific biased properties of dynamic systems. Simple equations combining derivatives and signals themselves are used. They allow to model bubbles and to investigate their dynamic behavior. A simple case study is added. It illustrates basic idea of method, which is proposed.\\
\\
Information processing and namely processing of information content in societies has a dynamic background. The latter is associated with gains and time delays which would lead to unstable behavior or to chaotic oscillations. In this paper, an attempt is met how to better understand rational/irrational  processes in societies.

\end{document}